\documentclass[amsmath,amssymb,11pt,superscriptaddress,reprint, preprintnumbers, notitlepage,aps,onecolumn,nofootinbib]{revtex4-1}
\pdfoutput=1 
\usepackage[utf8]{inputenc}
\usepackage[english]{babel}
\usepackage{amsmath}
\usepackage{graphicx}
\usepackage{dcolumn}
\usepackage{pbox}
\usepackage{amssymb}
\usepackage{epsfig}
\usepackage[dvipsnames]{xcolor}
\usepackage[normalem]{ulem}
\usepackage{slashed}
\usepackage{amssymb}
\usepackage{ mathrsfs }
\usepackage{color}
\usepackage[font=small]{caption}
\usepackage[font=small]{subcaption}

\usepackage{url}
\usepackage[makeroom]{cancel}
\usepackage{tikz}
\usetikzlibrary{decorations.pathmorphing}
\usepackage{enumitem} 
\usepackage{soul}
\usepackage{makecell}
\usepackage{pifont}
\usepackage{placeins}
\usepackage{tikz-feynman}
\usepackage{tabularray}
\UseTblrLibrary{booktabs}
\usepackage{booktabs,siunitx}
\usepackage{etoolbox}

\definecolor{MyDarkBlue}{rgb}{0.1, 0.1, 0.8} 
\definecolor{SBlue}{rgb}{0.2, 0.4, 0.7} 
\definecolor{MyLightBlue}{rgb}{0.22,0.51,0.9}
\definecolor{MyGreen}{rgb}{0.0, 0.5, 0.0}
\definecolor{BrickRed}{rgb}{0.8, 0.25, 0.33}
\RequirePackage{hyperref}
\hypersetup{colorlinks, citecolor=SBlue,linkcolor=MyDarkBlue, urlcolor=PineGreen}

\DeclareUnicodeCharacter{2212}{-}
\begin{document}
\title{\vspace{1cm}\Large 
$c\to u \gamma$ in  $J/\psi$, $B_c$ and $D_s$ decays: Standard Model and beyond}
\author{\bf Svjetlana Fajfer}
\email[E-mail:]{svjetlana.fajfer@ijs.si}
\affiliation{Jožef Stefan Institute, Jamova 39, P.\ O.\ Box 3000, SI-1001 Ljubljana, Slovenia}
\affiliation{Department of Physics, University of Ljubljana, Jadranska 19, 1000 Ljubljana, Slovenia}

\author{\bf Lara Kropej}
\email[E-mail:]{lara.kropej@merton.ox.ac.uk}
\affiliation{Jožef Stefan Institute, Jamova 39, P.\ O.\ Box 3000, SI-1001 Ljubljana, Slovenia}
\affiliation{Mathematical Institute, University of Oxford, Radcliffe Observatory Quarter, Woodstock Road, Oxford OX2 6GG, United Kingdom}

\begin{abstract}

Motivated by recent BESIII searches for the rare radiative decays $J/\psi\to \overline{D}^{0}\gamma$ and $D_s^+\to K^{*+}\gamma$, we investigate these processes within and beyond the Standard Model.  We consider$ J/\psi\to\overline{D}^{0}\gamma$, $B_c\to B_u^*\gamma$, and $D_s^+\to K^{*+}\gamma$, corresponding to charm, bottom, and strange spectator anti-quarks, respectively. A distinctive feature of our analysis is the comparison of the quark-level $c\to u\gamma$ transition in systems with different spectator quarks. This allows us to investigate how the short-distance $c\to u\gamma $ transition competes with long-distance hadronic contributions in three rather different environments. We determine their branching fractions using factorisation for the long-distance amplitudes and employ results of the recent determinations of the relevant hadronic form factors. Our focus is on both the long- and short-distance Standard Model contributions, and we subsequently investigate the maximal effects allowed by new physics, using existing constraints on the electromagnetic dipole coefficient $C_7$ derived from the measured branching ratio for $D^0 \to \rho^0 \gamma$. 
For the decay modes considered here, the vector-meson-dominance contribution is rather suppressed by cancellations among the $\rho$, $\omega$, and $\phi$ intermediate states. In contrast, the $D_s^+\to K^{+}\gamma$ decay is dominated by long-distance dynamics. Among the three channels studied, $B_c\to B_u^\gamma$ therefore offers the cleanest relative sensitivity to a possible new-physics enhancement of the $c\to u\gamma$ dipole transition.

\end{abstract}


\maketitle

\section{Introduction}

Flavour-changing neutral current (FCNC) processes are very important in the current studies of physics beyond the Standard Model (SM).
In particular, in these processes the GIM mechanism plays an important role, which is that  highly suppressed in the SM, and some extensions of the SM allow deviations from SM predictions \cite{Buras:2020xsm}. The largest effects are expected in the down quark transitions. On the other hand, among up-quark states, charm meson and baryon decays offer an opportunity to test the $c \to u $ transition.
These decays have been investigated for more than three decades \cite{Burdman:1995te,Fajfer:1998dv,deBoer:2017que,Gisbert:2020vjx}, offering possibilities to constrain new physics (NP) effects in the up sector. 
Recently, the BESIII Collaboration searched for the decay $J/\psi \to \overline{D}^0 \gamma$ and determined the upper limit $\mathcal{B}(J/\psi \to \overline{D}^0 \gamma)<9.1\times10^{-8}$ 
\cite{BESIII:2024kkx}. Some time ago, it was proposed to search for the $c \to u \gamma$ transition in the $B_c$ meson to $B_u^\ast \gamma$ \cite{Fajfer:1999dq,Losacco:2024zji}. In this case, the $\bar b$ anti-quark is the spectator, while in $J/\psi \to \overline{D}^0 \gamma$ the spectator is the $\bar c$ anti-quark. Our study includes the analysis of $D_s \to K^* \gamma$, where the spectator is a $\bar s$ anti-quark to compare it with the decay dynamics of $ J/\psi\to\overline{D}^{0}\gamma$, and  $B_c\to B_u^*\gamma$.

Motivated by the accumulated knowledge of $c \to u \gamma$ transitions from a theoretical point of view \cite{Burdman:1995te,Fajfer:1998dv,deBoer:2017que} and experimental searches at BESIII \cite{BESIII:2024zfv,BESIII:2026nvc}, Belle \cite{Abdesselam:2016yvr}, we first investigate the decay of $J/\psi \to \overline{D}^0\gamma$ within the SM and beyond. We then reconsider contributions and calculate rates for  $B_c\to B_u^*\gamma$ and $D_s^+\to K^{*+}\gamma$ using the same framework as in the case  $J/\psi\to \overline{D}^{0}\gamma$. 
The short-distance contribution (SD) is governed by the electromagnetic
dipole operator $O_7$, while four-quark current--current operators
generate long-distance radiative contributions  \cite{Burdman:1995te,Fajfer:1998dv}. Consequently, the long-distance contribution (LD) 
has to be understood quantitatively so that these modes are used
to constrain or eventually identify new physics in $C_7^{\rm eff}$.

 We consider 
long-distance contributions in all cases, assuming that the non-leptonic
transition $M\to M' V^0$, $V^0=\rho^0,\omega,\phi$, is followed by the vector-meson-dominance (VMD) conversion
$V^0\to\gamma $ \cite{Burdman:1995te,Fajfer:1998dv,deBoer:2017que,Gisbert:2020vjx}. In the case of $ J/\psi\to\overline{D}^{0}\gamma$ and $B_c\to B_u^*\gamma$,  this contribution is proportional to the effective coefficient
$a_2$ of the current-current operator $O_2$.
In all three decay modes, we include photon emissions from the initial and final states in our analysis. In the case of $B_c\to B_u^*\gamma$, it was found that the short-distance
 contribution from $c\to u\gamma$ can be of a similar size as the long-distance contribution even in the Standard Model  \cite{Fajfer:1999dq,Losacco:2024zji}.

The BESIII collaboration recently  \cite{BESIII:2026nvc}  determined the upper bound on the branching fraction 
$\mathcal{B}(D_s^+ \to K^{*+}(892) \gamma)<2.3\times10^{-4}$, which is not far above the predictions of \cite{Burdman:1995te,Fajfer:1998dv,deBoer:2017que,Cao:2023gfv}. 
However, in the case of $D_s^+\to K^{*+}\gamma$ both operators  $O_1$ and  $O_2$ contribute to the LD. Owing to the cancellation in the VMD combinations in the amplitude expressions
our numerical estimate makes long-distance contributions to the  $ J/\psi\to\overline{D}^{0}\gamma$ and $B_c\to B_u^*\gamma$ decay amplitudes either smaller or of the same order of magnitude as the Standard Model short-distance contribution proportional to $C_7^{\rm eff}$.
Thus, both $J/\psi\to\bar D^0\gamma$ and $B_c\to B_u^*\gamma$  
illustrate an important feature of rare charm radiative
transitions: they are potentially sensitive to new physics
in the $c\to u\gamma$ dipole operator. 

The main physics issues are introduced in Sec.~I, while the framework is discussed in Sec.~II. Sec.~III contains the study of the long- and short-distance contributions to $J/\psi \to \bar D^0 \gamma$. The update of the $B_c \to B_u^* \gamma$ decay amplitudes is described in Sec.~IV. Sec.~V contains the details of our analysis of the $D_s \to K^* \gamma$ decay amplitude.

\section{Framework: $c\to u \gamma $ transition in the SM }

We follow  the framework of weak effective  theory as presented in \cite{Adolph:2022ujd,Burdman:1995te,Fajfer:1998dv,deBoer:2017que} in which the  decay Lagrangians can be written in terms of dimension-six operators
\begin{align} 
   {\mathcal L}_{\text{eff}}^{\rm SCS} = \frac{4G_F}{\sqrt{2}}\left(\sum_{q=d,s}V_{cq}^* V_{uq}\sum_{i=1}^2 C_i O_i^{(qq)} + \sum_{i=3}^6 C_i O_i + \sum_{i=7}^8 \left(C_i O_i + C_i^\prime O_i^\prime\right)\right)\, ,
\label{Lag1}
\end{align}
\begin{align}
 {\mathcal L}_{\text{eff}}^{\rm CF} = \frac{4G_F}{\sqrt{2}} V_{cs}^* V_{ud}\sum_{i=1}^2 C_i O_i^{(ds)} \, , \quad 
 {\mathcal L}_{\text{eff}}^{\rm DCS} = \frac{4G_F}{\sqrt{2}} V_{cd}^* V_{us}\sum_{i=1}^2 C_i O_i^{(sd)}\,.
\end{align}
Here SCS, CF, and DCS stand for singly Cabibbo-suppressed, Cabibbo-favoured, and doubly Cabibbo-suppressed transitions, respectively; $G_F$ is the Fermi constant and $V_{ij}$ are elements of the Cabibbo--Kobayashi--Maskawa (CKM) matrix. 
In our study 
we will use the non-leptonic weak Lagrangian in the form 
\begin{equation} 
\mathcal{L}_W^{\Delta C=1}=\frac{G_F}{\sqrt{2}}V^*_{cq_j}V_{uq_i}[a_1  (\bar u \gamma^\mu P_L q_i) (\bar q_j \gamma_{\mu} P_L c)  
+ a_2 (\bar u\gamma^\mu P_L c) (\bar q_j \gamma_{\mu} P_L q_i)], 
\label{Lw}
\end{equation}
with $P_{L,R} = (1 \mp \gamma_5)/2$.  
We extend the SM operator basis of the four-quark operators $O_{1,2}^{(q q^\prime)}$ by allowing NP effects in the electromagnetic dipole operators $O_7$ and $O_7^{\prime}$  \cite{deBoer:2017que} ($q,q^\prime=d,s$)
\begin{equation}
  \begin{alignedat}{2}
  \label{Eq4}
    &O_7 = \frac{e \,m_c}{16\pi^2}  \left(\bar u _L \sigma^{\mu \nu} c_R\right)F_{\mu \nu} ,\enspace 
    &&O_7^\prime = \frac{e \,m_c}{16\pi^2}  \left(\bar u _R \sigma^{\mu \nu} c_L\right)F_{\mu \nu}\,.
  \end{alignedat}
\end{equation}

For comparison, Ref.~\cite{deBoer:2017que} uses another basis $\widetilde Q_1^{(q)}= (\bar u_\alpha\gamma^\mu P_Lc_\alpha) (\bar q_\beta\gamma_{\mu} P_L q_\beta)$, and $Q_2^{(q)} = \left(\bar u_L \gamma_\mu q_L\right) \left(\bar q_L \gamma^\mu c_L\right)$. We refer to their Wilson coefficients as $C_i^{\rm dBH}$. They depend on the scale at which they are determined. We set the scale to be equal to the charm quark mass.
Consequently, the coefficients in the basis of  Ref.\cite{Buras:2020xsm}  are $a_2=\frac12 C_1^{\rm dBH}$, and $a_1=C_2^{\rm dBH}-\frac{1}{2N_c}C_1^{\rm dBH}$, which results in, for $N_c=3$ and the leading-order ranges
$C_1^{\rm dBH}\in[-1.28,-0.83]$ and
$C_2^{\rm dBH}\in[1.14,1.06]$, this gives
\begin{equation}
 a_2 \in[-0.640,-0.415],
 \qquad
 a_1 \in[1.20,1.35].
 \label{eq:buras-current-current-coefficients}
\end{equation} 

\subsection{Short-distance contributions} 

The electromagnetic dipole operator relevant for the
$c\to u\gamma$ transition is
\begin{equation}
Q_7 =
\frac{e\,m_c}{16\pi^2}
\left(\bar u_L \sigma_{\mu\nu} c_R\right)F^{\mu\nu},
\qquad
Q_7' =
\frac{e\,m_c}{16\pi^2}
\left(\bar u_R \sigma_{\mu\nu} c_L\right)F^{\mu\nu}.
\end{equation}

The corresponding effective Wilson coefficient $C_7^{\rm eff}$, including two-loop
 QCD matrix elements of the operators $O_{3,...,6}$, determined at the charm quark mass, is found to be \cite{deBoer:2017que,Greub:1996wn} in the range
\begin{equation}
C_7^{\rm eff}
\in
\left[
-0.00151-(0.00556 i)_s +(0.00005 i)_{\rm CKM},
\,
-0.00088-(0.00327 i)_s +(0.00002 i)_{\rm CKM}
\right],
\label{Eq-C7}
\end{equation}
where the tiny weak CKM phase has been included, while $C_7^{\prime\,{\rm eff}} \simeq\frac{m_u}{m_c}\approx0$.
We use in our calculations $C_7^{\rm eff}=(-1.195 -4.380 \,i)
\times10^{-3}$.

\begin{figure}[htpb]
\centering
\includegraphics[width=0.6\textwidth]{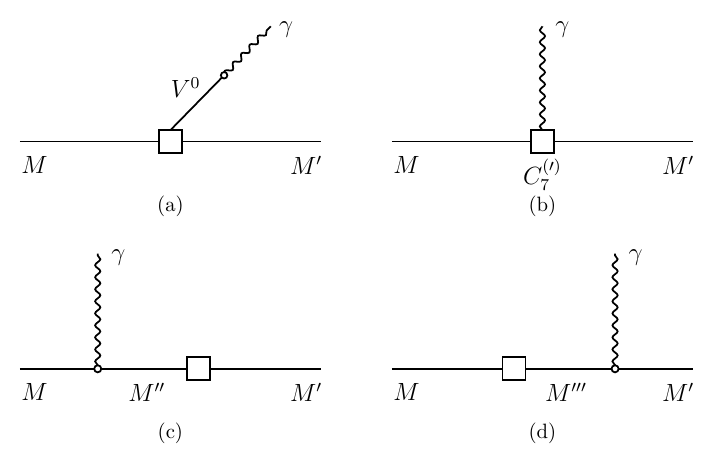}
\caption{Generic hadronic topologies for $M\to M'\gamma$: (a) a weak transition producing a neutral vector meson $V^0$, followed by $V^0\to\gamma$ conversion; (b) a local electromagnetic dipole transition proportional to $C_7^{(\prime)}$; (c) photon emission from the initial hadron through a virtual intermediate state $M''$; and (d) emission from the final hadron through a virtual state $M'''$. Squares denote weak vertices and circles denote electromagnetic vertices. Solid lines represent mesons and wavy lines denote the photon. 
}
\label{fig:radiative-topologies}
\end{figure}

The most interesting feature is that after including QCD running effects, the effective coefficient of the dipole operator  $C_7^\text{eff} $ is of order $ \mathcal{O} (10^{-3})$ \cite{deBoer:2017que}. 
Let us comment that the chromomagnetic operator $O_8$ contributes to
$D\to V\gamma$ decays both through its hadronic matrix
elements and indirectly through renormalisation-group
mixing into the electromagnetic dipole operator $O_7$  \cite{deBoer:2017que,Greub:1996wn,Lyon:2012fk}.
Consequently, new-physics contributions to $C_8$ can
modify the effective coefficient $C_7$ and induce
sizeable direct $CP$ asymmetries in radiative charm
decays~\cite{Lyon:2012fk}. SD contribution (dipole contribution is presented in Fig.~\ref{fig:radiative-topologies} (b). 

\subsection{Long-distance contributions}

The calculation of the long-distance contributions is based on the Lagrangian in Eq.(\ref{Lag1}) and applies to hadronic interactions that may be present in $M\to M' \gamma$ decays as illustrated in 
Fig.~\ref{fig:radiative-topologies}. 
 Long-distance contributions are given in Fig.~\ref{fig:radiative-topologies} as (a) the VMD contribution and  (c) and (d), which present photon emission from the initial and final states, in which the annihilation diagrams come from Eq.~(\ref{Lag1}).

We assume  "naive factorisation" of the hadronic amplitudes. The non-leptonic
transition $M\to M' V^0$, $V^0=\rho^0,\omega,\phi$, is followed by the VMD conversion
$V^0\to\gamma $ \cite{Burdman:1995te,Fajfer:1998dv,deBoer:2017que,Gisbert:2020vjx}.
In the following, we briefly repeat the basic features of the VMD. 

\subsubsection{Vector Meson Dominance}
The electromagnetic interaction with the quark current is
${\mathcal L}_{em} = e A_\mu  \, j^\mu =  A_\mu  \,  (\frac{2}{3} \bar u \gamma^\mu  u - \frac{1}{3} \bar d \gamma^\mu d -\frac{1}{3} \bar s \gamma^\mu s) $.
The vector-meson dominance means that each quark current is replaced by the vector mesons which have appropriate quark structure:
$ \bar u \gamma^\mu  u  \to (\frac{1}{{\sqrt 2}} (\rho^{0 \mu}+\omega^\mu)$,  
$\bar d \gamma^\mu  d  \to \frac{1}{{\sqrt 2}} (-\rho^{0 \mu}+\omega^\mu)$,  
$ \bar s \gamma^\mu s \to \phi^\mu) $.  
The Lagrangian has dimension four, and therefore the VMD Lagrangian is 
\begin{equation}
   {\mathcal L}_{em} = e \,A_\mu  \frac{1}{{\sqrt 2}}( g_\rho \rho^{0 \mu} + \frac{1}{3} g_\omega \omega^\mu - \frac{{\sqrt 2}}{3}  g_\phi \phi^\mu ).
\label{VMD2}
\end{equation}
In our calculation, we use the factor $g_V$ 
which is defined as $<0| V^\mu | V(q,\epsilon_V) > = g_v \epsilon_{V}^\mu $, with $V^\mu = \rho^{0\mu}, \omega^\mu, \phi^\mu$.  One $g_V$ arises from the matrix element $ <V(q,\epsilon_V| \bar q \gamma_\mu q| 0> $ while the second one from the VMD Lagrangian in Eq. (\ref{VMD2}).
Using $V_{cs}V_{us}^*\simeq -V_{cd}V_{ud}^*$
the three vector-meson contributions are governed by the parameter 
\begin{equation}
G_{\rm VMD}
=
-\frac{g_\rho^2}{2m_\rho^2}
+\frac{g_\omega^2}{6m_\omega^2}
+\frac{g_\phi^2}{3m_\phi^2}.
\label{eq:GVMD}
\end{equation}
The Lagrangian in Eq.~(\ref{VMD2}) is not by itself gauge invariant, as pointed out in Refs.~\cite{Golowich:1994zr,Cao:2023gfv}; a fully gauge-invariant VMD construction is discussed in Ref.~\cite{Soares:1995ri}. In the amplitudes below we therefore impose electromagnetic gauge invariance on the full radiative amplitude. The resulting expressions should be understood as a phenomenological gauge-invariant implementation of VMD rather than as a derivation from Eq.~(\ref{VMD2}) alone. 

 This simplified picture effectively describes the hadronic dynamics in $c\to u \gamma$ transitions. Other effects might be present, as considered in the final-state interactions in Ref. \cite{Cao:2023gfv} or in the VMD contribution; the parameter $G_{\rm VMD}$ introduces uncertainty due to cancellations among parameters. The vector meson couplings are well established by experiments \cite{ParticleDataGroup:2026} and lattice QCD \cite{Aoki:2021kgd}. 
We use the normalisation
$\langle 0 | J_{\rm em}^{\mu} | V(q,\epsilon) \rangle
= g_V \epsilon_V^\mu$.

\begin{table}[htbp]
\centering
\begin{tabular}{|c|c|c|}
\hline
Meson ($V$) & $g_V$ latt. QCD $[\mathrm{GeV}^2]$&  $g_V$ exp. (PDG) $[\rm{GeV^2]}$ \\
\hline\hline
$\rho(770)$   & $0.167 \pm 0.005$ & $0.165 \pm 0.003$ \\
$\omega(782)$ & $0.152 \pm 0.008$ & $0.145 \pm 0.004$ \\
$\phi(1020)$  & $0.232 \pm 0.004$ & $0.229 \pm 0.002$ \\
\hline
\end{tabular}
\caption{Vector meson couplings $g_V$ calculated by lattice QCD
\cite{Aoki:2021kgd} and extracted from experimental data for $V\to \ell ^+ \ell^-$
\cite{Workman:2022ynf} for the $\rho$, $\omega$, and $\phi$ mesons.}
\label{gV-LE}
\end{table}
Using Eq.~(\ref{eq:GVMD}) and Table~\ref{gV-LE}, we obtain
\begin{equation}
G_{\rm VMD}^{\rm exp}=(-0.109\pm0.930)\times10^{-3}\,\mathrm{GeV}^2
\label{eq:GVMDII}
\end{equation}
For comparison, the lattice inputs give $G_{\rm VMD}^{\rm lat}=(0.348\pm1.650)\times10^{-3}\,\mathrm{GeV}^2$.
The experimental central contributions from $\rho$, $\omega$ and $\phi$ are
$(-22.6487,\,5.7207,\,16.8193)\times10^{-3}\,\mathrm{GeV}^2$.
The errors are obtained by first-order propagation of the independent coupling uncertainties in Table~\ref{gV-LE}; vector-meson mass uncertainties are neglected.
We use the unrounded combinations $G_{\rm VMD}^{\rm exp}=-1.09\times10^{-4}\,\mathrm{GeV}^2$ and $G_{\rm VMD}^{\rm lat}=3.48\times10^{-4}\,\mathrm{GeV}^2$.



\subsubsection{ Form factors for hadronic matrix elements}

To calculate amplitudes for the $M\to M'\gamma$ transitions based on naive factorisation, we need to use matrix elements of the weak currents. They are parametrised in terms of form factors. A commonly used parametrisation of the vector-current matrix element is \cite{Wirbel:1985ji}

\begin{equation}
\langle V(k,\varepsilon)|
\bar q \gamma_\mu Q
|P(p)\rangle
=
\frac{2iV(q^2)}{m_P+m_V}
\epsilon_{\mu\nu\rho\sigma}
\varepsilon^{*\nu}p^\rho k^\sigma
.
\label{V-current}
\end{equation}
The axial-vector current matrix element is parametrised as
\begin{align}
\langle V(k,\varepsilon)|
\bar q\gamma_\mu\gamma_5 Q
|P(p)\rangle
={}&
2m_V A_0(q^2)
\frac{\varepsilon^*\cdot q}{q^2}q_\mu
+ (m_P+m_V)A_1(q^2)
\left(
\varepsilon_\mu^*
-
\frac{\varepsilon^*\cdot q}{q^2}q_\mu
\right)
\nonumber\\
&-\frac{A_2(q^2)\,\varepsilon^*\cdot q}{m_P+m_V}
\left[
(p+k)_\mu
-
\frac{m_P^2-m_V^2}{q^2}q_\mu
\right].
\label{A-current}
\end{align}
The absence of a spurious pole at \(q^2=0\) requires 
$A_0(0) =\frac{m_P+m_V}{2m_V}A_1(0)- \frac{m_P-m_V}{2m_V}A_2(0)$. 
For the tensor current, one writes \cite{Ball:2004rg}
\begin{equation}
\langle V(k,\varepsilon)|
\bar q\sigma_{\mu\nu}q^\nu Q
|P(p)\rangle
=
2iT_1(q^2)
\epsilon_{\mu\nu\rho\sigma}
\varepsilon^{*\nu}p^\rho k^\sigma.
\end{equation}

For the pseudotensor current,
\begin{align}
\langle V(k,\varepsilon)|
\bar q\sigma_{\mu\nu}\gamma_5q^\nu Q
|P(p)\rangle
={}&
T_2(q^2)
\left[
(m_P^2-m_V^2)\varepsilon_\mu^*
-
(\varepsilon^*\cdot q)(p+k)_\mu
\right]
\nonumber\\
&+
T_3(q^2)(\varepsilon^*\cdot q)
\left[
q_\mu
-
\frac{q^2}{m_P^2-m_V^2}(p+k)_\mu
\right].
\end{align}


At \(q^2=0\), one has the kinematic relation $T_1(0)=T_2(0)$.

\subsection{Beyond the Standard Model}

As in previous studies of radiative charm decays
\cite{Burdman:1995te,Fajfer:1998dv,deBoer:2017que,Biswas:2017eyn},
new physics might modify the electromagnetic dipole operators and,
consequently, the Wilson coefficients $C_7^{(\prime)\rm eff}$. In the Standard Model
$C_7^{\prime\,\rm eff}\simeq0$.
The parity-conserving and parity-violating short-dostance amplitudes are then
\begin{equation}
A_{\rm PC}^{\rm SD}=  \frac{G_F e\,m_c}{2\sqrt{2}\pi^2}\,
\left[(C_7^{\rm eff}+\delta C_7) + (C_7^{\prime\,\rm eff}+\delta C_7')\right] \,T^{M \,M'}(0),
\enspace
A_{\rm PV}^{\rm SD}
=
\frac{G_F e\, m_c}{2\sqrt{2}\pi^2}\,
\left[(C_7^{\rm eff}+\delta C_7) - (C_7^{\prime\,\rm eff}+\delta C_7')\right] \,T^{M \,M'}(0).
\label{eq:JpsiC7amps}
\end{equation}

Among the measured branching ratios  \cite{ParticleDataGroup:2026}, the smallest error is in the case of $D^0 \to \rho^0 \gamma$. The branching ratio 
$\mathcal{B} (D^0 \to \rho^0 \gamma) = (1.81 \pm 0.31)\times 10^{-5}$ enables us to set the limit on the maximal value of the short-distance Wilson coefficient $|C_7^{\rm{eff}}|_{\rm{max}}$ \cite{deBoer:2018buv}. In the Appendix we use the same  framework as in the rest of the calculations to derive the  maximally allowed $|C_7^{\rm{eff}}|_{\rm{max}}$ from the branching ratio for
$D^0 \to \rho^0 \gamma$. We reproduce the commonly used conservative estimate $|C_7^{(\prime)\rm{eff}}|_{\rm{max}}\lesssim0.3$ of Ref.~\cite{deBoer:2018buv}. This should not be interpreted as a model-independent experimental bound: the measured $D^0\to\rho^0\gamma$ amplitude contains long-distance contributions and can interfere with the dipole amplitude.
We use $|C_7^{\rm{eff}}|_{\rm{max}}=0.3$ below only as a representative benchmark for potentially present new physics.

\section{The $J/\psi \to \bar D^0 \gamma$  amplitude}

We apply the above-described strategy to the decay $J/\psi \to \bar D^0 \gamma$ amplitude. 
The most general decomposition of the $J/\psi \to \bar D^0 \gamma$ amplitude contains the parity-conserving (${\cal A}_{PC}$) and parity-violating (${\cal A}_{PV}$) amplitudes. The  gauge-invariant form is  as usual 
\begin{equation}
{\cal M}(J/\psi(p,\epsilon_\psi)  \to\bar D^0(p-q) \gamma (q,\epsilon_\gamma)) =
e\,\frac{G_F}{\sqrt2}\,
V_{cd}V_{ud}^*\,a_2\,
\epsilon_\psi^\mu\epsilon_\gamma^{*\nu}
\Big[
i\,{\cal A}_{PC}\,
\epsilon_{\mu\nu\alpha\beta}p^\alpha q^\beta
+{\cal A}_{PV}
\left(
g_{\mu\nu}\,p\cdot q-p_\nu q_\mu
\right)
\Big].
\label{eq:JpsiDgamma}
\end{equation}
The corresponding decay width, after averaging over the three
$J/\psi$ polarisations, is
\begin{equation}
\Gamma(J/\psi\to\bar D^0\gamma)
=
\frac{M_{J/\psi}^3}{96\, \pi}
\left(
1-\frac{m_D^2}{M_{J/\psi}^2}
\right)^3
\frac{e^2G_F^2}{2}
|V_{cd}V_{ud}^*|^2\, a_2^2
\left(
|{\cal A}_{PC}|^2+
|{\cal A}_{PV}|^2
\right).
\label{eq:GammaJpsiDgamma}
\end{equation}
No charge-conjugate multiplicity is included here.

\subsection{Long-distance  contribution to 
$J/\psi\to\bar D^0\gamma$}

First, we consider the long-distance (LD)  contribution from the diagram (a) in Fig. 1, referred to as the VMD contribution. 

\paragraph{Vector-meson-dominance contribution to $J/\psi\to\bar D^0\gamma$.}

The radiative decay $J/\psi\to\bar D^0\gamma$, with
$\bar D^0=(\bar c u)$, receives a long-distance contribution from
current-current operators through intermediate neutral vector mesons.
Following the VMD treatment of radiative charm decays
\cite{Burdman:1995te,Fajfer:1998dv,deBoer:2017que}, we consider
$J/\psi\to\bar D^0 V^0$, with $V^0=\rho^0,\omega,\phi$, followed by the transition
$V^0\to\gamma$.

In the simple "naive factorisation" approach, the relevant colour-suppressed contribution to the non-leptonic decay amplitude 
$J/\psi\to\bar D^0 V^0$ (using $\langle V^0|(\bar q q)_V|0\rangle  =g_V \epsilon_V^\mu$) is:
\begin{equation}
{\cal M}(J/\psi\to\bar D^0 V^0)
=
i \frac{G_F}{\sqrt2}\,a_2
\sum_{q=d,s}V_{cq}V_{uq}^*
\langle\bar D^0|(\bar u c)_{V-A}|J/\psi\rangle
\langle V^0|(\bar q q)_V|0\rangle ,
\label{eq:JpsiDV}
\end{equation}
and it has to be combined with the VMD interaction, when $V^0$ transforms to a photon, following Eq.(\ref{VMD2}). 
The VMD contribution to the parity-conserving amplitude is easily found
\begin{equation}
{\cal A}_{\rm{PC}}^{\rm VMD}
=
\frac{2V(0)}
{M_{J/\psi}+m_D}\,
G_{\rm VMD}.
\label{eq:APC}
\end{equation}
The gauge invariance holds for this contribution.
The parity-violating contribution is obtained from the axial-vector
form factors after imposing electromagnetic gauge invariance.
Here we need $g_V(0)$. We assume that $g_V(0) $ do not differ very much from the value $g_V(m_V^2)$ with $V=\rho^0,\, \omega,\, \phi$. 
After conversion of the neutral vector meson into a real photon,
terms proportional to $q_\mu$ do not contribute because
$q\cdot\epsilon_\gamma=0$. The remaining axial contribution is
therefore proportional to
$(M_{J/\psi}+m_D) A_1(0) (\epsilon\cdot\epsilon_\gamma^*) $. This expression alone is not manifestly gauge invariant. To satisfy  electromagnetic gauge
invariance of the amplitude, the physical amplitude can be written in terms of 
 the transverse tensor \break $ [(\epsilon\cdot q)(\epsilon_\gamma^*\cdot p) - (\epsilon\cdot\epsilon_\gamma^*)(p\cdot q)]$. Since $(p\cdot q)
=(M_{J/\psi}^2-m_D^2)/2$, this requires that 
matching the coefficient of
$(\epsilon\cdot\epsilon_\gamma^*)$ leads to a condition
$-(p\cdot q)\,{\cal A}_{\rm{PV}}^{\rm VMD} =
(M_{J/\psi}+m_D)A_1(0)\,G_{\rm VMD}$,
and hence
\begin{equation}
{\cal A}_{\rm PV}^{\rm VMD}
=
-\frac{2A_1(0)}
{M_{J/\psi}-m_D}\,
G_{\rm VMD}.
\label{eq:APV}
\end{equation}

The overall sign depends on the convention used for the axial current
and for the definition of ${\cal A}_{PV}$; it has no effect on the
decay rate in the absence of interference with another amplitude. 

\paragraph{Annihilation pole contributions to $J/\psi\to\bar D^0\gamma$.}
\label{sec:JpsiAdditionalPoles}

The transitions $J/\psi\to P\gamma$, followed by weak annihilation $P\to\bar D^0$, contribute through virtual $P=\eta,\eta'$ poles, as in Fig.~\ref{fig:radiative-topologies}(c). 
We introduce
 $f_\eta^q=f_q\cos\phi$, $f_\eta^s=-f_s\sin\phi$,
$ f_{\eta'}^q=f_q\sin\phi$, $f_{\eta'}^s=f_s\cos\phi$ ~\cite{Feldmann:1998vh}.
The individual $d$-quark axial-current decay constant is $f_P^d=f_P^q/\sqrt2$ and we denote  $\lambda_q=V_{cq}^*V_{uq}$.

In color-suppressed factorization and the quark-flavor mixing scheme~\cite{Feldmann:1998vh}, their amplitudes are
\begin{equation}
 {\cal M}_P=A_P\,\epsilon_{\mu\nu\alpha\beta}
 \epsilon_\psi^\mu\epsilon_\gamma^{*\nu}p^\alpha q^\beta,
 \qquad A_{PV}^{P}=0,
\end{equation}
with
\begin{align}
 A_\eta&=-\frac{eG_F a_2 f_D m_D^2\,g_{\psi\eta\gamma}}
 {\sqrt2\,(m_D^2-m_\eta^2)}
 \left(\frac{\lambda_d f_q\cos\phi}{\sqrt2}
                  -\lambda_s f_s\sin\phi\right),
 \label{eq:JpsiEtaPole}\\
 A_{\eta'}&=-\frac{eG_F a_2 f_D m_D^2\,g_{\psi\eta'\gamma}}
 {\sqrt2\,(m_D^2-m_{\eta'}^2)}
 \left(\frac{\lambda_d f_q\sin\phi}{\sqrt2}
                  +\lambda_s f_s\cos\phi\right).
 \label{eq:JpsiEtapPole}
\end{align}
The factor $1/\sqrt2$ in the $d$-quark term follows from the normalized $(u\bar u+d\bar d)/\sqrt2$ flavor state. The radiative couplings are defined by ${\cal M}(J/\psi\to P\gamma)=e\,g_{\psi P\gamma}\epsilon_{\mu\nu\alpha\beta}\epsilon_\psi^\mu\epsilon_\gamma^{*\nu}p^\alpha q^\beta$ and extracted from
\begin{equation}
 \Gamma(J/\psi\to P\gamma)=\frac{\alpha}{24}
 |g_{\psi P\gamma}|^2\frac{(M_{J/\psi}^2-m_P^2)^3}{M_{J/\psi}^3}.
\end{equation}
We use the measured branching fractions $\mathcal B(J/\psi\to\eta\gamma)=1.085\times10^{-3}$ and $\mathcal B(J/\psi\to\eta'\gamma)=5.250\times10^{-3}$~\cite{ParticleDataGroup:2026}, $\Gamma_{J/\psi}=92.600~\mathrm{keV}$, $a_2=-0.5$, $f_D=0.212~\mathrm{GeV}$, $f_q=1.07\,f_\pi$, $f_s=1.34 \, f_\pi$, $f_\pi=130.2~\mathrm{MeV}$ and $\phi=39.3^\circ$ \cite{Feldmann:1998vh}. Neglecting $\lambda_b$, we take $\lambda_d=-\lambda_s=-0.218$. We determine $|g_{\psi \eta \gamma}| =4.14 \times 10^{-3}$ ${\rm GeV^{-1}}$ and $|g_{\psi \eta' \gamma}| =8.53 \times 10^{-3}$ ${\rm GeV^{-1}}$.
The virtual pseudoscalar carries momentum $p_D$. Equations~(\ref{eq:JpsiEtaPole})--(\ref{eq:JpsiEtapPole}) assume a factorised weak transition at $p_D^2=m_D^2$ and approximate the radiative couplings there by their on-shell values. This off-shell continuation and the relative phases are not fixed by the measured radiative rates.

The individual pole branching fractions follow directly from
\begin{equation}
 \mathcal B_P(J/\psi\to\bar D^0\gamma)
 =\frac{(M_{J/\psi}^2-m_D^2)^3}
 {96\,\pi\, M_{J/\psi}^3\Gamma_{J/\psi}}\,|A_P|^2,
 \qquad P=\eta,\eta',
 \label{eq:JpsiFullCoefficientWidth}
\end{equation}
and are $ \mathcal B_\eta=4.710\times10^{-19}$,
$ \mathcal B_{\eta'}=6.525\times10^{-19}$.


\paragraph{LD sum and comparison with SD.}
After matching all terms to the same tensor convention, the calculated LD subset has
\begin{align}
 A_{PC}^{\rm LD}&=A_{PC}^{\rm VMD}
       +e^{i\delta_\eta}|A_\eta|+e^{i\delta_{\eta'}}|A_{\eta'}|,\nonumber\\
 A_{PV}^{\rm LD}&=A_{PV}^{\rm VMD}.
 \label{eq:JpsiPartialLDsum}
\end{align}
Then, the VMD coefficients are
\begin{equation}
 A_{PC}^{\rm VMD}=-i \,\frac{eG_F}{\sqrt2}V_{cd}V_{ud}^*a_2\,\mathcal A_{PC}^{\rm VMD},\qquad
 A_{PV}^{\rm VMD}= i \,\frac{eG_F}{\sqrt2}V_{cd}V_{ud}^*a_2\,\mathcal A_{PV}^{\rm VMD},\qquad
\end{equation}
The pole phases in Eq.~(\ref{eq:JpsiPartialLDsum}) include the corresponding tensor-convention phases; only their magnitudes enter the ranges. For the experimental central $G_{\rm VMD}$, these magnitudes are $5.724\times10^{-12}$ and $1.829\times10^{-11}~\mathrm{GeV}^{-1}$, respectively.
We choose a constructive sum of both pole PC amplitudes and the VMD PC amplitude. This gives
\begin{equation}
 \mathcal B_{\rm LD} =  \mathcal B_{\mathrm{VMD}}+  \mathcal B_{\eta}+   \mathcal B_{\eta'}=2.617\times10^{-18}.
 \label{eq:JpsiPartialLDreference}
\end{equation}
This phase choice is illustrative; it is not fixed by factorisation or the radiative branching fractions. Varying the two pole phases independently at central $G_{\rm VMD}$ gives $9.282\times10^{-20}\leq\mathcal B_{\rm LD}\leq2.617\times10^{-18}$. Scanning also the quoted experimental $G_{\rm VMD}$ interval gives
 $1.343\times10^{-20}\leq\mathcal B_{\rm LD}\leq1.418\times10^{-17}$.
These are sensitivity ranges within the stated pole model. 
\subsection{Short-distance contribution for $J/\psi\to\bar D^0\gamma$  }

The short-distance contribution to
$J/\psi\to\bar D^0\gamma$ is induced by the electromagnetic dipole
operators given in Eq. (\ref{Eq4}). We use the effective coefficient $C_7^{\rm eff}$ given in Eq.(\ref{Eq-C7}) \cite{deBoer:2017que,deBoer:2018buv}.  
In this convention the CKM factors are already contained in
$C_7^{\rm eff}$. 
The tensor form factors are defined such that at $q^2=0$ is $T_1(0)=T_2(0)\equiv T(0)$.
The parity-conserving and parity-violating short-distance amplitudes in the SM 
are then ${\cal A}_{PC}^{\rm{SD}}= \frac{G_F \, e\, m_c}{2\sqrt{2}\,\pi^2}\left(C_7^{\rm eff}+C_7^{\prime\,{\rm eff}}\right)
T(0)$,  and ${\cal A}_{PV}^{\rm{SD}}
= \frac{G_F\,e \, m_c}{2\sqrt{2}\pi^2} \left(C_7^{\rm eff}-C_7^{\prime\,{\rm eff}}\right)
T(0)$. Since $C_7^{\prime\,{\rm eff}}\simeq0$, one has
\begin{equation}
\left|{\cal A}_{PC}^{\rm{SD}}\right|
=
\left|{\cal A}_{PV}^{\rm{SD}}\right|
=
\frac{e\,G_F\, m_c}{2\sqrt{2}\,\pi^2}
|C_7^{\rm eff}|\,T(0).
\end{equation}

The decay width, including the average over the three
$J/\psi$ polarisations, is 
\begin{equation}
\Gamma_{\rm{SD}}(J/\psi\to\bar D^0\gamma)
=
\frac{M_{J/\psi}^3}{96\,\pi}
\left(
1-\frac{m_D^2}{M_{J/\psi}^2}
\right)^3
\left(
|{\cal A}_{PC}^{\rm{SD}}|^2+
|{\cal A}_{PV}^{\rm{SD}}|^2
\right).
\end{equation}

A physical total rate requires the coherent addition of the long- and short-distance contributions. The separate LD and SD rates cannot determine this interference without their relative phases and the missing LD amplitudes.
The common SM input is $C_7^{\rm eff}=(-1.195 -4.380 i)
\times10^{-3}$.
The form factors of Ref.~\cite{Galkin:2026kdf} are $V(0)=0.479$ and $A_1(0)=0.380$.
We quote one channel transition $c \to u \gamma$ in all contributions to  $J/\psi\to\bar D^0\gamma$. Neglecting direct CP asymmetry, the charge-conjugate sum is twice this rate. The amplitudes are
${\cal A}_{PC}^{\rm VMD}
=(-2.098\pm17.949)\times10^{-5}\,\mathrm{GeV}$,
${\cal A}_{PV}^{\rm VMD}
=(6.703\pm57.345)\times10^{-5}\,\mathrm{GeV}$. The corresponding decay rate is 
${\cal B}_{\rm VMD}^{\rm exp}=(1.01\pm17.31)\times10^{-19}$.
The displayed symmetric errors propagate only $G_{\rm VMD}$ to first order.
Scanning its quoted input interval gives the sensitivity range $0\leq {\cal B}_{\rm VMD}^{\rm exp}\leq9.23\times10^{-18}$. 
For the lattice input we obtain ${\cal A}_{PC}^{\rm VMD,lat}=(6.723\pm31.857)\times10^{-5}\,\mathrm{GeV}$ and ${\cal A}_{PV}^{\rm VMD,lat}=(-2.148\pm10.178)\times10^{-4}\,\mathrm{GeV}$,
${\cal B}_{\rm VMD}^{\rm lat}=(1.04\pm9.84)\times10^{-18}$.
The corresponding lattice-input sensitivity range  is $0\leq {\cal B}_{\rm VMD}^{\rm lat}\leq3.42\times10^{-17}$.

With the estimate based on heavy-quark and large-recoil symmetry relations \cite{Charles:1998dr,Beneke:2000wa} $T_1(q^2)= \frac{m_{J/\psi}}{m_{J/\psi}+m_D}\,
V(q^2)$ giving $T_1(0)=T_2(0)\equiv T^{J/\psi D}(0)\simeq0.3$. 
With $T(0)=0.3$, we find
${\cal B}_{\rm SD}=1.28\times10^{-12}|C_7^{\rm eff}|^2$ and
${\cal B}_{\rm SD}^{\rm SM}=2.64\times10^{-17}$.
The central VMD contribution is much smaller than the SM dipole contribution, but the VMD input uncertainty and the unquantified tensor-form-factor uncertainty limit the significance of this hierarchy.



Numerically, for the maximal value of $|C_7^{\rm{eff}}|_{\rm{max}}=0.3$ the decay width is  $\Gamma(J/\psi\to \bar D^0\gamma)_{\rm BSM} \lesssim
1.067\times10^{-17}~{\rm GeV}$,
and using $\Gamma_{J/\psi}=92.6~{\rm keV}$, we get for the branching ratio
\begin{equation}
{\cal B}(J/\psi\to \bar D^0\gamma)_{\rm BSM}
\lesssim
1.15\times10^{-13}.
\end{equation}

For a complex $C_7^{\rm eff}$ and, in general, $C_7'$, a physical total rate requires coherent addition 
\begin{equation}
 \mathcal B_{\rm tot}=\frac{(M_{J/\psi}^2-m_D^2)^3}
 {96\,\pi\, M_{J/\psi}^3\Gamma_{J/\psi}}
 \left(|A_{PC}^{\rm SD}+A_{PC}^{\rm LD}|^2
       +|A_{PV}^{\rm SD}+A_{PV}^{\rm LD}|^2\right).
\end{equation}

\paragraph{Charge-conjugate channel and the NP normalisation.}
Neglecting direct CP violation and adding $J/\psi\to D^0\gamma$, we get $2.3\times10^{-13}$. Taking both $|C_7|=|C_7'|=0.3$ would independently double the dipole rate and is a different benchmark.
Note that the coherent total branching ratio requires information about relative phases, beyond the separate LD and SD branching fractions.
The Table 
\ref{tab:BRcomparison} gives single-channel  entries. The experimental upper limit for the branching fraction is  $9.1\times10^{-8}$ and the $J/\psi$ point in Fig.~\ref{fig:transitions} refer to the charge-conjugate sum.

\begin{table}[htbp]
\centering
\begin{tabular}{|lc|}
\hline
Contribution & Branching fraction \\
\hline\hline
VMD   & $1.011\times10^{-19}$ \\
$\eta$ pole   & $4.71\times10^{-19}$ \\
$\eta'$ pole   & $6.53\times10^{-19}$ \\
Coherent LD sum & $2.62\times10^{-18}$ \\
SD, SM  & $2.64 \times10^{-17}$ \\
\hline
\end{tabular}
\caption{Branching fractions for the separate contributions to one charge
channel $J/\psi\to\bar D^0\gamma$. The LD sum uses the constructive sum 
VMD+$\eta$+$\eta'$  with experimental $G_{\rm VMD}$. The relative phases within the LD parity-conserving amplitude are chosen to be zero, so VMD, \(\eta\), and \(\eta'\) add constructively.
The relative LD–SD phase is not specified.
}
\label{tab:JpsiContributions}
\end{table}

\section{Long- and short-distance contributions to $B_c^+\to B_u^{*+}\gamma$}
\label{sec:BcLDinterference}
The $\bar b$ spectator makes this decay a probe of $c\to u\gamma$~\cite{Fajfer:1999dq,Losacco:2024zji}. We include light-vector VMD, the lowest weak-annihilation poles, and the electromagnetic dipole contribution. All coefficients below contain their weak and electromagnetic prefactors. Our convention is
\begin{equation}
 {\cal M}(B_c^+ (p) \to B_u^{*+}(p-q, \epsilon') \, \gamma(q,\epsilon))=i\epsilon_\mu^{\prime *}\epsilon_\nu^*
 \left[A_{\rm PV}(p^\nu q^\mu-g^{\mu\nu}p\cdot q)
 +iA_{\rm PC}\epsilon^{\mu\nu\alpha\beta}p'_\alpha p_\beta\right],
 \label{eq:BcAmp}
\end{equation}
where $p'=p-q$ and the vector state has a polarization $\epsilon'$ and photon carries index $\epsilon$.

The VMD contributions for $B_c^+\to B_u^{*+}\gamma$  are
\begin{equation}
 A_{\rm PC}^{\rm VMD}=\frac{eG_F}{\sqrt2}V_{cs}^*V_{us}a_2
 \frac{2G_{\rm VMD}V^{B_cB_u^*}(0)}{m_{B_c}+m_{B_u^*}},\quad
 A_{\rm PV}^{\rm VMD}=\frac{eG_F}{\sqrt2}V_{cs}^*V_{us}a_2
 \frac{2G_{\rm VMD}A_1^{B_cB_u^*}(0)}{m_{B_c}-m_{B_u^*}}.
 \label{eq:BcVMDamps}
\end{equation}
In the pole model of Ref.~\cite{Fajfer:1999dq}, weak annihilation gives
\begin{align}
 A_{\rm PC}^{B_c^*}&=-\frac{eG_F}{\sqrt2}a_1 \, V_{cb}^*V_{ub}
 \frac{\mu_{B_c}g_{B_c^*}g_{B_u^*}}{m_{B_c*}^2 -m_{B_u*}^2},\quad
 A_{\rm PC}^{B_u}=-\frac{eG_F}{\sqrt2}a_1V_{cb}^*V_{ub}
 \frac{\mu_{B_u}m_{B_c}^2 f_{B_c}f_{B_u}}
 {m_{B_c}^2-m_{B_u}^2}\,,\nonumber\\
 A_{\rm PV}^{B_c^*}&=A_{\rm PV}^{B_u}=0.
 \label{eq:BcAnnCombined}
\end{align}
Note that both amplitudes have an  additional phase factor $e^{i\delta_{\rm ann}}$. Here the coupling $\mu_{B_{c(u)}}$ describe full radiative vertices of the initial (final) state radiaton. 
We retain $V_{cb}^*V_{ub}=|V_{cb}^*V_{ub}|e^{-i\gamma}$. 

For $C_7'=0$ and negligible $m_u$, the dipole amplitudes are
\begin{align}
 A_{\rm PC}^{\rm SD}&=-\frac{eG_Fm_cF_1(0)}{4\sqrt2\pi^2}C_7^{\rm eff}\quad
 A_{\rm PV}^{\rm SD}=-\frac{eG_Fm_cF_2(0)}{2\sqrt2\pi^2}C_7^{\rm eff},
 \qquad F_1(0)=2T_1(0),\quad F_2(0)=T_2(0).
 \label{eq:BcC7amps}
\end{align}
The complex coefficient $C_7^{\rm eff}$ already includes CKM suppression~\cite{deBoer:2017que}. 
The coherent coefficients and rate are
\begin{align}
 A_{\rm PC}^{\rm LD}&=A_{\rm PC}^{\rm VMD}+A_{\rm PC}^{B_c^*}+A_{\rm PC}^{B_u},\quad
 A_{\rm PV}^{\rm LD}=A_{\rm PV}^{\rm VMD},\nonumber\\
 \Gamma_{\rm tot}&=\frac{1}{4\pi}
 \left(\frac{m_{B_c}^2-m_{B_u^*}^2}{2m_{B_c}}\right)^3
 \left(|A_{\rm PC}^{\rm LD}+A_{\rm PC}^{\rm SD}|^2
      +|A_{\rm PV}^{\rm LD}+A_{\rm PV}^{\rm SD}|^2\right).
 \label{eq:BcCoherentWidth}
\end{align}
Setting SD or LD to zero defines their separate widths. Their interference is obtained by replacing the final parentheses by $2\operatorname{Re}[A_{\rm PC}^{\rm SD}(A_{\rm PC}^{\rm LD})^*+A_{\rm PV}^{\rm SD}(A_{\rm PV}^{\rm LD})^*]$.
For $B_c^-$, conjugate the weak CKM factors, including those within $C_7^{\rm eff}$, but retain the strong phases. 

\paragraph{Inputs and branching fractions}

We use $m_{B_c}=6.274$ GeV, $m_{B_u^*}=5.325$  GeV,, $m_{B_u}=5.280$  GeV, and $m_c=1.250$GeV, with $\tau_{B_c}=0.510$ ps. The updated $B_c^*$ pole mass is $m_{B_c^*}=6.339$ GeV from ATLAS~\cite{ATLAS:2026BcStarObservation}. The transition form factors are $V(0)=0.090\pm0.020$, $A_1(0)=-0.170\pm0.030$ and $T_1(0)=T_2(0)=0.230\pm0.040$~\cite{Aliev:2006vs}.
For a consistent recent set of decay constants, we use the continuum-extrapolated CLQCD preprint~\cite{Cai:2026BDecayConstants}: $f_{B_c}=(0.451\pm 0.0052)$ GeV, $f_{B_u}=(0.1898\pm0.0035)$ GeV, $f_{B_c^*}=(0.438\pm0.0043)$ GeV and $f_{B_u^*}=(0.1812\pm 0.0042)$ GeV. The light-quark results are used in the isospin approximation. Converting with $g_{B^*}=m_{B^*}f_{B^*}$ gives the inputs
\begin{align}
& f_{B_c}=0.451~\mathrm{GeV},\quad f_{B_u}=0.190~\mathrm{GeV}, \quad 
g_{B_c^*}=(2.776\pm0.027)~\mathrm{GeV}^2,\quad
 g_{B_u^*}=(0.965\pm0.022)~\mathrm{GeV}^2,\nonumber\\
 &|\mu_{B_c}|=(298.500\pm7.200)\times10^{-3}~\mathrm{GeV}^{-1},\quad
 |\mu_{B_u}|=1.440^{+0.260}_{-0.270}~\mathrm{GeV}^{-1}.
 \label{eq:BcPoleCouplings}
\end{align}
The couplings $ |\mu_{B_c}|$ and $|\mu_{B_u}|$ are from the published light-front calculation of Ref.~\cite{Wang:2025voj}
{and the charged-$B$ NLO light-cone sum rule of Ref.~\cite{Pullin:2021ebn}, respectively. Both match our normalization $\Gamma(V\to P\gamma)=\alpha|\mu_{VP}|^2(m_V^2-m_P^2)^3/(24\,m_V^3)$. 
We take $|V_{cs}^*V_{us}|=0.218$, $|V_{cb}^*V_{ub}|=1.4\times10^{-4}$, $\gamma=65.7^\circ$, $\delta_{\rm ann}=0$, and the experimental $G_{\rm VMD}$ of Eq.~(\ref{eq:GVMDII}). 

For the interference, we use the complex midpoint in Eq.~(7) of Ref.~\cite{deBoer:2017que},
$C_7^{\rm eff}=(-1.195-4.380i)\times10^{-3}$ ($|C_7^{\rm eff}| =4.53\times10^{-5} $). 
We retain the pole-residue signs appearing in our amplitude
definitions and set the additional strong phase
$\delta_{\rm ann}=0$. These signs and this phase choice
define our reference prescription.
\begin{table}[htbp]
\centering
\begin{tabular}{|lc|}
\hline
Contribution & Branching fraction \\
\hline\hline
VMD   & $4.704\times10^{-12}$ \\
Annihilation   & $1.481\times10^{-9}$ \\
Coherent LD sum & $1.483\times10^{-9}$ \\
SD & $2.275\times10^{-9}$ \\
SD--LD interference & $1.965\times10^{-9}$ \\
Coherent SM total & $5.723\times10^{-9}$ \\
\hline
\end{tabular}
\caption{Branching fractions for the separate contributions
to $B_c^+\to B_u^{*+}\gamma$ at the reference pole-residue
signs, $\delta_{\rm ann}=0$, and $\gamma=65.7^\circ$, with
$V_{cb}^*V_{ub}=|V_{cb}^*V_{ub}|e^{-i\gamma}$.
The SD contribution is evaluated for 
$C_7^{\rm eff}=(-1.195-4.380i)\times10^{-3}$.
The coherent SM total includes the SD--LD interference
contribution $+1.965\times10^{-9}$.}

\label{tab:BcLDinterference}
\end{table}
For $|C_7^{\rm eff}|_{\rm max}=0.3$, $C_7'=0$, the separate NP dipole rate is $(9.935\pm3.456)\times10^{-6}$; its displayed uncertainty is from $T(0)$ only. 
These sensitivities exclude unknown strong phases, higher poles, off-shell residues and factorization uncertainties.

The newer CLQCD constants \cite{Cai:2026CLQCD} differ from published HPQCD values~\cite{Colquhoun:2015BcDecayConstants} and the precise charged-$B$ result of Ref.~\cite{Bazavov:2018BDecayConstants}. Using $f_{B_c}=(0.427\pm 0.006)$ GeV, $f_{B_u}=(0.189 \pm10.004)$ GeV, $f_{B_c^*}=(0.422\pm10.003)$ GeV and $f_{B_u^*}=(0.175.0\pm0.006)$ GeV instead, with the other updated inputs fixed, gives $\mathcal B_{\rm LD}=1.3\times10^{-9}$ and $\mathcal B_{\rm tot}=5.4\times10^{-9}$; this is an alternative input scenario, not a statistical error. Recent QCD sum-rule results~\cite{CostaiReina:2026BcDecayConstants} provide a further comparison. 
Within this updated reference model, the LD rate is below but comparable to the SM dipole rate. Their coherent sum depends on both weak and strong phases; CKM suppression alone does not make the pole contribution negligible relative to the cancelled VMD term.

 \section{Updating long- and short-distance contributions to $D_s^+\to K^{*+}\gamma$}
\label{sec:DsKstgamma}
We use the factorized pole-plus-VMD description of Ref.~\cite{deBoer:2017que}, including photon emission from both hadron legs. The full amplitudes are defined by
\begin{align}
 {\cal M}(D_s (p)^+\to K^{*+}(k,\epsilon_{K*} )\gamma(q,\epsilon_\gamma) =&\epsilon_{\mu\nu\alpha\beta}q^\mu\epsilon_\gamma^{*\nu}
 p^\alpha\epsilon_{K^*}^{*\beta}A_{\rm PC}
 +i\left[(\epsilon_{K^*}^*\!\cdot q)(\epsilon_\gamma^*\!\cdot p')
 -(p'\!\cdot q)(\epsilon_{K^*}^*\!\cdot\epsilon_\gamma^*)\right]A_{\rm PV},
  \label{eq:DsKst_amp}
\end{align}
with $p'=p-q$. Both structures are transverse; all weak and electromagnetic factors are included in $A_{\rm PC,PV}$.

\subsection{Long-distance amplitudes}

We present the long-distance amplitudes with all additional
strong phases set to zero, retaining the signs displayed
below and the signed experimental $G_{\rm VMD}$.
Following the contributions presented in Fig. (\ref{fig:radiative-topologies}), (c) and (d), 
the initial $D_s^*$ and final $K^+$  magnetic poles give
\begin{align}
 A_{\rm PC}^{D_s^*}&=\frac{eG_F}{\sqrt2}V_{cs}^*V_{us}a_1
 \frac{|\mu_{D_s^*D_s}|\,m_{D_s^*}f_{D_s^*}m_{K^*}f_{K^*}}
 {m_{D_s^*}^2-m_{K^*}^2}\,,\quad 
 A_{\rm PC}^{K}=\frac{eG_F}{\sqrt2}V_{cs}^*V_{us}a_1
 \frac{|\mu_{K^*K}|\,m_{D_s}^2f_{D_s}f_K}
 {m_{D_s}^2-m_K^2}\,.
 \label{eq:DsMagneticPoles}
\end{align}
Their PV coefficients vanish in this approximation. The gauge-invariant annihilation diagram contribution, including photon emission from $K^*$, and the axial $\phi$ conversion are
\begin{align}
 A_{\rm PV}^{\rm el}&=\frac{e \,G_F}{\sqrt2}V_{cs}^*V_{us}
 \frac{ 2\,a_1\,m_{K^*}f_{D_s}f_{K^*}}{m_{D_s}^2-m_{K^*}^2}\,,\quad 
 A_{\rm PV}^{\phi}=\frac{e\,G_F}{\sqrt2}V_{cs}^*V_{us}
 \frac{2a_1(m_{D_s}+m_\phi)m_{K^*}f_{K^*}f_\phi}
 {3m_\phi(m_{D_s}^2-m_{K^*}^2)}
 A_1^{D_s\phi}(m_{K^*}^2)\,.
 \label{eq:DsElectricPhi}
\end{align}
 Note that factor $-1/3$ is from the electromagnetic current, the $s$-quark charge. The remaining VMD penguin amplitudes are
\begin{align}
 A_{\rm PC}^{\rm penguin}&=\frac{eG_F}{\sqrt2}V_{cs}^*V_{us}
 \frac{2a_2G_{\rm VMD}V^{D_sK^*}(0)}{m_{D_s}+m_{K^*}}\,,\quad 
 A_{\rm PV}^{\rm penguin}=\frac{eG_F}{\sqrt2}V_{cs}^*V_{us}
 \frac{2a_2G_{\rm VMD}A_1^{D_sK^*}(0)}{m_{D_s}-m_{K^*}}.
 \label{eq:DsPenguin}
\end{align}
Thus
\begin{align}
 A_{\rm PC}^{\rm LD}&=A_{\rm PC}^{D_s^*}+A_{\rm PC}^{K}+A_{\rm PC}^{\rm penguin},\quad 
 A_{\rm PV}^{\rm LD} =A_{\rm PV}^{\rm el}+A_{\rm PV}^{\phi}+A_{\rm PV}^{\rm penguin}.
 \label{eq:DsHybridSum}
\end{align}
The cancellation in $G_{\rm VMD}$ affects the penguin terms only.


The magnetic-coupling magnitudes follow from
 $\Gamma(V\to P\gamma)=\frac{\alpha}{3}|\mu_{VP}|^2
 \left(\frac{m_V^2-m_P^2}{2m_V}\right)^3$. We use $\Gamma(D_s^*\to D_s\gamma)=(54.9 \pm 5.4)$ eV~\cite{Meng:2024DsStarRadiative}, $\Gamma(K^{*+}\to K^+\gamma)=(50.0\pm 5.0)$ keV~\cite{PDG:2025KstarInputs}, 
 $f_{D_s^*}=(0.274\pm 0.006)$ GeV~\cite{Donald:2014DsStarWidth}, $f_{D_s}=(0.2499\pm 0.0004)$ GeV~\cite{Bazavov:2018BDecayConstants}, and $f_K=0.1557$ GeV \cite{PDG:2025KstarInputs}. 
 On-shell radiative couplings are used as constants in our calculations.  
To account small $SU(3)$ flavour symmetry breaking we retain  $g_{K^*}=\frac{m_{K^*}}{m_\rho}g_\rho=0.190~\mathrm{GeV}^2$,
$f_{K^*}=\frac{g_\rho}{m_\rho}$, $ f_\phi=\frac{g_\phi}{m_\phi}$ 
following \cite{Fajfer:1998dv}.
Our inputs are 
$a_1=1.2$, $a_2=-0.5$, $V^{D_sK^*}(0)=1.04$, $A_1^{D_sK^*}(0)=0.57$ and $A_1^{D_s\phi}(m_{K^*}^2)=0.675$~\cite{Melikhov:2000yu,Fajfer:1998dv}.
$|\mu_{D_s* D_s}|= 0.09172$ GeV$^{-1}$, $|\mu_{K* K}|= 0.83399$ GeV$^{-1}$. Masses and decay widths are taken from\cite{ParticleDataGroup:2026}. 
For the central results, we set all additional LD strong
phases to zero and retain the signed experimental
$G_{\rm VMD}$. 
The relative strong phases are not determined by the
radiative widths.
We vary the six LD phases while keeping the input
magnitudes fixed, with no restriction on one phase
imposed by others' values.
It gives $\mathcal B_{\rm LD}\leq (4.74\pm0.24)\times10^{-5}$. This range illustrates the phase dependence and is not
a statistical uncertainty. Further uncertainties arise
from the form factors, factorisation, $SU(3)$ breaking,
off-shell effects and omitted intermediate states.
\subsection{Short- distance contribution}
For $C_7'=0$, the full dipole amplitudes are
\begin{equation}
 A_{\rm PC}^{\rm SD}=A_{\rm PV}^{\rm SD}
 =\frac{eG_Fm_cT^{D_sK^*}(0)}{2\sqrt2\pi^2}C_7^{\rm eff},
 \qquad.
 \label{eq:DsDipoleComplex}
\end{equation}
We use $T^{D_sK^*}(0)=0.710$ from Ref.\cite{Melikhov:2000yu} and $ m_c=1.270~\mathrm{GeV}$. 
With the full LD coefficients above, the coherent width is
\begin{equation}
 \Gamma_{\rm tot}=\frac{1}{4\pi}
 \left(\frac{m_{D_s}^2-m_{K^*}^2}{2m_{D_s}}\right)^3
 \left(|A_{\rm PC}^{\rm LD}+A_{\rm PC}^{\rm SD}|^2
      +|A_{\rm PV}^{\rm LD}+A_{\rm PV}^{\rm SD}|^2\right).
 \label{eq:DsHybridInterference}
\end{equation}
The separate dipole branching fraction is $\mathcal B_{\rm SD}=7.59\times10^{-4}|C_7^{\rm eff}|^2$ or $\mathcal B_{\rm SD}=(1.67\pm0.81)\times10^{-8}$. The central value and error are the midpoint and half-width of the branching-fraction interval generated by the quoted $C_7^{\rm eff}$ range. The tensor-form-factor uncertainty remains additional and unquantified.

\begin{table}[htbp]
\centering
\begin{tabular}{|lc|}
\hline
Contribution & Branching fraction \\
\hline\hline
LD & $(4.74\pm0.2)\times10^{-5}$ \\
SD & $(1.67\pm0.81)\times10^{-8}$ \\
SD--LD interference & $-4.50\times10^{-7}$ \\
Coherent SM total & $4.72\times10^{-5}$ \\
SD, NP benchmark & $6.83\times10^{-5}$ \\
Experiment & $<2.3\times10^{-4}$ \\
\hline
\end{tabular}
\caption{Branching fractions for $D_s^+\to K^{*+}\gamma$ with
$a_1=1.2$, $a_2=-0.5$, experimental $G_{\rm VMD}$,
and all additional LD strong phases set to zero.
The SM entries use $C_7^{\rm eff}=(-1.195-4.380i)\times10^{-3}$.
The interference is a signed contribution to the branching fraction.
The NP entry is the separate dipole rate for
$|C_7^{\rm eff}|=0.3$ and $C_7'=0$.}
\label{tab:DsContributions}
\end{table}
At fixed zero additional strong phases, we find 
$\mathcal B_{\rm tot}^{\rm SM}\leq (4.70 \pm0.24 )\times10^{-5}$, with error coming from the input parameters and the error coming from dipole interaction at the per cent level. 

\section{Conclusions}

Motivated by the recent BESIII results on the radiative charm decays $J/\psi \to \bar D^0 \gamma$ and $D_s^+ \to K^{*+} \gamma$, we reconsider $c\to u\gamma$ transitions in $J/\psi \to \bar D^0 \gamma$, $B_c \to B_u^* \gamma$, and $D_s^+ \to K^{*+} \gamma$. 
We identified two types of long-distance contributions: weak annihilation and the vector meson dominance model. In our calculations, we relied on relevant parameters from lattice QCD as input, along with experimental data. The uncertainties of the VMD contributions are dominated by the uncertainty in $G_{\rm VMD}$, which is strongly enhanced by cancellations among the $\rho$, $\omega$, and $\phi$ contributions. In particular, for $J/\psi\to\bar D^0\gamma$, the resulting rate uncertainty is highly non-Gaussian; the same is true for the $B_c$ experimental VMD input. For the $B_c\to B_u^*\gamma$ dipole contribution, the quoted form actor uncertainty originates from the tensor form factors. 

We present our main results in Table \ref{tab:BRcomparison}. For $J/\psi\to\bar D^0\gamma$, the VMD contribution is strongly suppressed by cancellations in $G_{\rm VMD}$; the corrected central VMD branching fraction is $1.011\times10^{-19}$. Including the $\eta,\eta'$ poles in the factorization and constant-vertex model gives  the rate $2.617\times10^{-18}$, below the SM dipole range $(1.454-4.174)\times10^{-17}$. Unknown relative phases, off-shell residues and additional weak poles prevent us for this decay mode from establishing a precise hierarchy for the full LD contribution. The representative benchmark $|C_7^{\rm eff}|=0.3$, derived by requiring that the dipole contribution not exceed the measured $D^0\to\rho^0\gamma$ rate,  
leads to the branching ratio being more than five orders of magnitude smaller than the current upper bound for $\mathcal{B}(J/\psi \to \bar D^0 \gamma)$. In the case of $B_c \to B_u^* \gamma$, the Standard Model dipole contribution dominates the central VMD term alone. Once the lowest-pole weak-annihilation contribution is included, the combined LD and SD rates are comparable; their coherent sum depends on the CKM weak phase and the relative strong phases, as discussed in Sec.~\ref{sec:BcLDinterference}. Finally, in the case of $D_s^+ \to K^{*+} \gamma$, the short-distance contribution is suppressed in
comparison with the long-distance one. The $D_s^+ \to K^{*+} \gamma$ is therefore in the same category as the rest of the $D$ mesons, in which the long-distance contributions dominate. 
\begin{table}[h]
\centering
\setlength{\tabcolsep}{4pt}
\begin{tabular}{|l|cccc|}
\hline
Decay & $\mathrm{BR}_{\rm LD}$ & $\mathrm{BR}_{\rm SD}^{\rm SM}$ & $\mathrm{BR}_{\rm SD}^{\rm NP}$ & $\mathrm{BR}_{\rm exp}$ \\
\hline\hline
$J/\psi\to\bar D^0\gamma$ & $2.617\times10^{-18}$ & $(2.814\pm1.360)\times10^{-17}$ & $1.152\times10^{-13}$ & $<9.1\times10^{-8}\ $ 
\\[1mm]
$B_c^+\to B_u^{*+}\gamma$ & $1.483\times10^{-9}$ & $(2.426\pm1.173)\times10^{-9}$ & $9.935\times10^{-6}$ & $-$ \\[1mm]
$D_s^+\to K^{*+}\gamma$ & $4.74\times10^{-5}$ & $(1.67\pm0.81)\times10^{-8}$ & $6.833\times10^{-5}$ & $<2.3\times10^{-4}$ \\
\hline
\end{tabular}
\caption{Comparison of separate LD and dipole branching fractions using the experimental $G_{\rm VMD}$ input. The $J/\psi$ LD entry is the constructive VMD+$\eta$+$\eta'$ sum of Sec.~\ref{sec:JpsiAdditionalPoles}, with central $G_{\rm VMD}$ and both pole PC amplitudes aligned with VMD. 
The $B_c^+$ LD entry includes coherent VMD and lowest-pole annihilation at $\gamma=65.7^\circ$ and $\delta_{\rm ann}=0$, using the updated lattice, magnetic-coupling and mass inputs of Sec.~\ref{sec:BcLDinterference}.  The $D_s^+$ LD entry results from initial and final magnetic poles, annihilation with a photon emission and the $\phi$ VMD contributions. 
SM SD $\pm$ values denote the midpoint of the $C_7^{\rm eff}$ range, with form factor uncertainties additional. The NP benchmark uses $|C_7^{\rm eff}|=0.3$ and $C_7'=0$.
LD and SD columns exclude mutual interference and must not be added to obtain a prediction of the total rate; coherent-sum results are given in the text. The $J/\psi$ theoretical entries refer to one charge channel, whereas its experimental limit includes $\gamma D^0+\mathrm{c.c.}$; double the theoretical entries for comparison with this sum.}
\label{tab:BRcomparison}
\end{table}

\begin{figure}[h]
\centering
\includegraphics[width=1.2\textwidth]{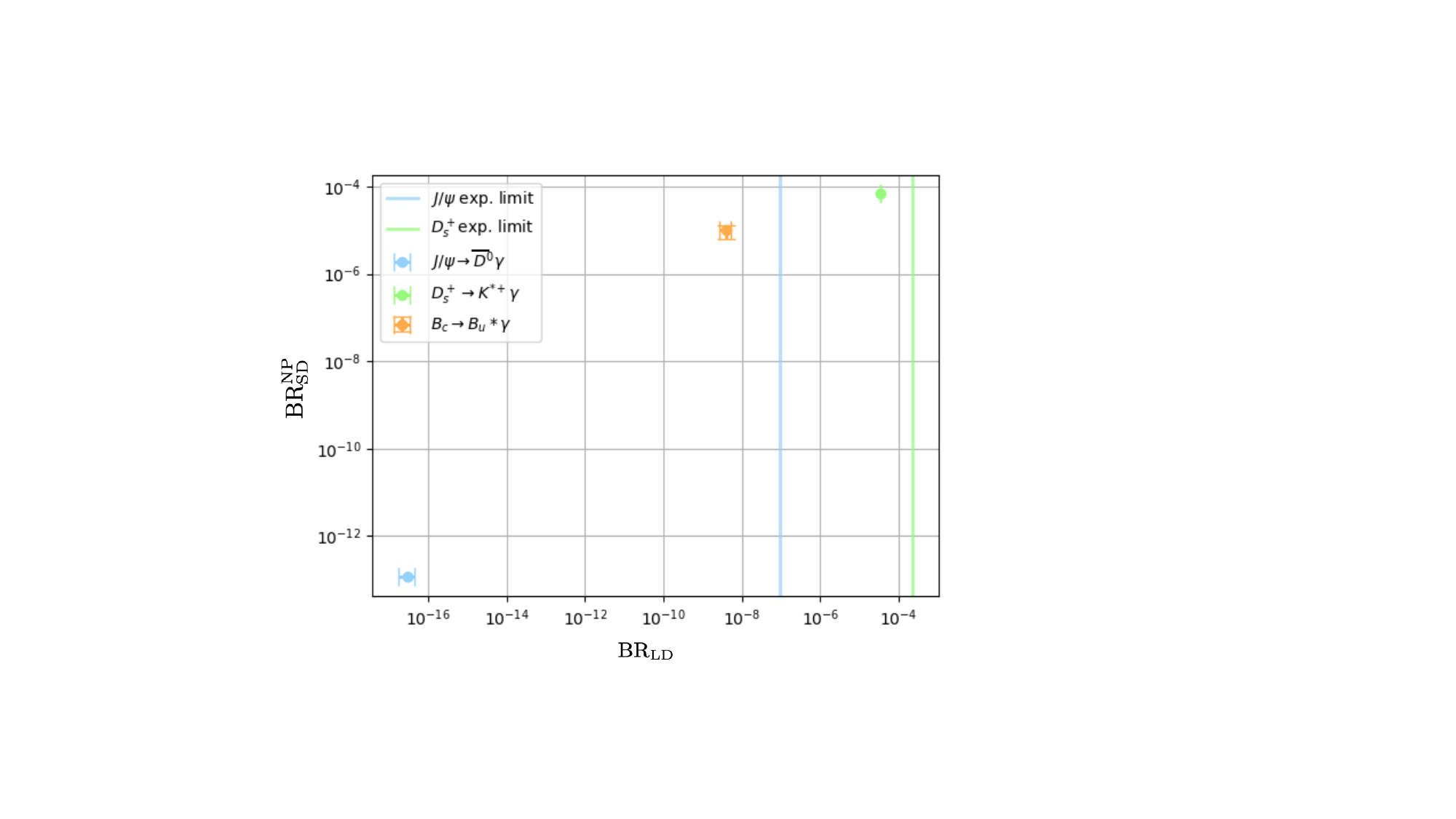}
\caption{LD reference branching fractions versus the separate NP dipole contribution for $|C_7^{\rm eff}|_{\rm max}=0.3$ and $C_7'=0$, updated to the calculations summarized in Table~\ref{tab:BRcomparison}. 
The vertical lines mark experimental upper limits on total branching fractions and provide reference scales, not separate bounds on LD in the presence of }
\label{fig:transitions}
\end{figure}
It should be stressed that the branching fractions quoted in
Table~\ref{tab:BRcomparison} refer to exclusive ground-state
transitions. 
In the annihilation diagram, excited states might also contribute. 
However, their decay dynamics are not so well known, and it exceeds the scope of this approach. 
Experimentally, however, such states may produce feed-down
backgrounds if the additional soft photons or pions are not
reconstructed. This issue is particularly relevant for
$B_c\to B_u^*\gamma$, since the subsequent decay
$B_u^*\to B_u\gamma_{\rm soft}$
is an intrinsic part of the experimental signature. Thus, the observable final state contains a hard photon from the primary
$B_c\to B_u^*\gamma$ transition and a second, considerably softer
photon from the $B_u^*$ decay.

\subsection{Acknowledgments}
SF and LK acknowledge the financial support from the Slovenian Research Agency (research core funding No. P1-0035 and N1-0321).

\appendix

\section{$D^0\to\rho^0\gamma$}
\label{sec:D0rhogammaLD}
The Belle measurement of $D^0\to\rho^0\gamma$~\cite{Belle:2016mtj} provides a useful scale for dipole contributions, but its interpretation requires LD amplitudes beyond the GIM-suppressed neutral-vector penguin. We distinguish that VMD component from the magnetic-pole and spectator-type contributions in the description of Refs.~\cite{Fajfer:1998dv,deBoer:2017que}.


Appendix~C, Eq.~(62), of Ref.~\cite{deBoer:2017que} contains an initial $D^{*0}$ magnetic pole, a light-pseudoscalar pole contribution and a spectator-type vector-conversion amplitude.
The hybrid calculations in Refs.~\cite{Fajfer:1998dv,deBoer:2017que}, including pole and spectator contributions and their phase variations, give branching fractions of characteristic order
\begin{equation}
 {\cal B}_{\rm LD}^{\rm SM}(D^0\to\rho^0\gamma)\sim10^{-6}-10^{-5}.
 \label{eq:DrhoLDLiteratureScale}
\end{equation}

\paragraph{Short-distance contribution and interference}
For the electromagnetic dipole operator, with $C_7'=0$, the full coefficients are~\cite{deBoer:2017que}
\begin{equation}
 A_{\rm PC}^{\rm SD}=A_{\rm PV}^{\rm SD} =\frac{1}{\sqrt2}   \frac{e\,G_F \,m_c}{2\sqrt2 \pi^2}
 C_7^{\rm eff}T^{D\rho}(0).
 \label{eq:DrhoSDcoefficient}
\end{equation}
The explicit $1/\sqrt2$ is the neutral-$\rho$ flavour factor in this tensor-form-factor convention. The coefficient $C_7^{\rm eff}$ already includes CKM suppression. Taking $T_1^{D\rho}(0)=T_2^{D\rho}(0)=T^{D\rho}(0)=0.66$ from Ref.~\cite{Melikhov:2000yu}, $m_c=1.27$ GeV and $\tau_{D^0}=0.4101$ ps gives
\begin{equation}
 A_{\rm PC}^{\rm SD}=A_{\rm PV}^{\rm SD}
 =7.50\times10^{-8}C_7^{\rm eff}~\mathrm{GeV}^{-1},
 \qquad
 {\cal B}_{\rm SD}=2.56\times10^{-4}|C_7^{\rm eff}|^2.
 \label{eq:DrhoSDbranching}
\end{equation}
The SM coefficient-magnitude range used in the text gives ${\cal B}_{\rm SD}^{\rm SM}\simeq(2.91-8.34)\times10^{-9}$.
Note  that, a complex $C_7^{\rm eff}$ must be combined with the complex LD amplitudes before taking absolute squares. The rate alone does not determine the relative strong phases.

\subsection{Interpretation of the dipole-saturation benchmark}

Belle measured ${\cal B}(D^0\to\rho^0\gamma)=(1.77\pm0.30\pm0.07)\times10^{-5}$~\cite{Belle:2016mtj}. If the dipole contribution is required to account for at most the observed rate, with cancellations against LD excluded by assumption, Eq.~(\ref{eq:DrhoSDbranching}) gives $|C_7^{\rm eff}|\lesssim0.26$ at the experimental central value. The approximate one-standard-deviation upper rate, $2.08\times10^{-5}$, gives $|C_7^{\rm eff}|\lesssim0.29$.

These estimates motivate the illustrative value used throughout the comparison,
\begin{equation}
 |C_7^{\rm eff}|_{\rm{max}}=0.3\qquad\text{(dipole-saturation benchmark)},
 \label{eq:DrhoDipoleBenchmark}
\end{equation}
consistent with the value derived in Ref.~\cite{deBoer:2018buv}. This prescription is not a phase-independent upper bound on the Wilson coefficient. The additional LD amplitudes in Eq.~(\ref{eq:DrhoLDLiteratureScale})
can interfere destructively with the dipole contribution, permitting values that a dipole-only comparison would exclude. A quantitative constraint requires their magnitudes, relative phases and hadronic uncertainties, together with the tensor-form-factor uncertainty, that is presently  not possible to account.

\bibliography{main}

@article{Fajfer:1998dv,
    author = "Fajfer, S. and Prelovsek, Sasa and Singer, P.",
    title = "{Long distance contributions in D ---{\ensuremath{>}} V gamma decays}",
    eprint = "hep-ph/9801279",
    archivePrefix = "arXiv",
    reportNumber = "IJS-TP-98-01, TECHNION-PH-98-01",
    doi = "10.1007/s100520050356",
    journal = "Eur. Phys. J. C",
    volume = "6",
    pages = "471--476",
    year = "1999"
}

@article{Burdman:1995te,
    author = "Burdman, Gustavo and Golowich, Eugene and Hewett, JoAnne L. and Pakvasa, Sandip",
    title = "{Radiative weak decays of charm mesons}",
    eprint = "hep-ph/9502329",
    archivePrefix = "arXiv",
    reportNumber = "SLAC-PUB-6692, SLAC-PUB-95-6692, FERMILAB-PUB-94-412-T, UMHEP-415, UH-511-811-94",
    doi = "10.1103/PhysRevD.52.6383",
    journal = "Phys. Rev. D",
    volume = "52",
    pages = "6383--6399",
    year = "1995"
}

@article{BESIII:2024kkx,
    author = "Ablikim, Medina and others",
    collaboration = "BESIII",
    title = "{Search for the rare decay J/{\ensuremath{\psi}}{\textrightarrow}{\ensuremath{\gamma}}D0+c.c. at BESIII}",
    eprint = "2408.08826",
    archivePrefix = "arXiv",
    primaryClass = "hep-ex",
    doi = "10.1103/PhysRevD.110.112012",
    journal = "Phys. Rev. D",
    volume = "110",
    number = "11",
    pages = "112012",
    year = "2024"
}

@article{deBoer:2017que,
    author = "de Boer, Stefan and Hiller, Gudrun",
    title = "{Rare radiative charm decays within the standard model and beyond}",
    eprint = "1701.06392",
    archivePrefix = "arXiv",
    primaryClass = "hep-ph",
    reportNumber = "DO-TH-16-20, DO-TH 16/20, QFET-2016-18",
    doi = "10.1007/JHEP08(2017)091",
    journal = "JHEP",
    volume = "08",
    pages = "091",
    year = "2017"
}

@book{Buras:2020xsm,
    author = "Buras, Andrzej",
    title = "{Gauge Theory of Weak Decays}",
    doi = "10.1017/9781139524100",
    isbn = "978-1-139-52410-0, 978-1-107-03403-7",
    publisher = "Cambridge University Press",
    month = "6",
    year = "2020"
}

@article{Fajfer:1999dq,
    author = "Fajfer, S. and Prelovsek, Sasa and Singer, P.",
    title = "{FCNC transitions $c \to u \gamma$ and $s \to d \gamma$ in $B_c \to B_{u}^* \gamma$ and $B_s \to B_{d}^* \gamma$ decays}",
    eprint = "hep-ph/9901252",
    archivePrefix = "arXiv",
    reportNumber = "IJS-TP-99-01, TECHNION-PH-99-02",
    doi = "10.1103/PhysRevD.59.114003",
    journal = "Phys. Rev. D",
    volume = "59",
    pages = "114003",
    year = "1999",
    note = "[Erratum: Phys.Rev.D 64, 099903 (2001)]"
}

@article{Adolph:2022ujd,
    author = "Adolph, Nico and Hiller, Gudrun",
    title = "{Rare radiative decays of charm baryons}",
    eprint = "2203.14982",
    archivePrefix = "arXiv",
    primaryClass = "hep-ph",
    reportNumber = "DO-TH 21/26",
    doi = "10.1103/PhysRevD.105.116001",
    journal = "Phys. Rev. D",
    volume = "105",
    number = "11",
    pages = "116001",
    year = "2022"
}

@article{deBoer:2018buv,
  author         = {de Boer, Sjoerd and Hiller, Gudrun},
  title          = {Flavor and new physics opportunities with rare charm decays into leptons},
  journal        = {Phys. Rev. D},
  volume         = {98},
  number         = {3},
  pages          = {035041},
  year           = {2018},
  doi            = {10.1103/PhysRevD.98.035041},
  eprint         = {1805.08516},
  archivePrefix  = {arXiv},
  primaryClass   = {hep-ph}
}

@article{Abdesselam:2016yvr,
  author         = {Abdesselam, A. and others and {Belle Collaboration}},
  title          = {Measurement of the branching ratio and $CP$ asymmetry of $D^0 \to \rho^0 \gamma$ and search for the decays $D^0 \to \phi \gamma$ and $D^0 \to \overline{K}^{*0}\gamma$},
  journal        = {Phys. Rev. Lett.},
  volume         = {118},
  number         = {5},
  pages          = {051801},
  year           = {2017},
  doi            = {10.1103/PhysRevLett.118.051801},
  eprint         = {1603.03257},
  archivePrefix  = {arXiv},
  primaryClass   = {hep-ex}
}

@article{Galkin:2026kdf,
    author = "Galkin, V. O. and Sukhanov, I. S.",
    title = "{Exclusive semileptonic and nonleptonic J/{\ensuremath{\psi}} decays}",
    eprint = "2604.01782",
    archivePrefix = "arXiv",
    primaryClass = "hep-ph",
    doi = "10.1103/46pk-nmsc",
    journal = "Phys. Rev. D",
    volume = "113",
    number = "11",
    pages = "113008",
    year = "2026"
}

@article{BESIII:2026nvc,
    author = "Ablikim, Medina and others",
    collaboration = "BESIII",
    title = "{Search for the radiative decay $ {D}_s^{+}\to \gamma {K}^{\ast }{(892)}^{+} $}",
    eprint = "2601.16476",
    archivePrefix = "arXiv",
    primaryClass = "hep-ex",
    doi = "10.1007/JHEP07(2026)266",
    journal = "JHEP",
    volume = "07",
    pages = "266",
    year = "2026"
}

@article{BESIII:2024zfv,
    author = "Ablikim, Medina and others",
    collaboration = "BESIII",
    title = "{Search for the radiative decays D$^{+}${\textrightarrow} {\ensuremath{\gamma}}{\ensuremath{\rho}}$^{+}$ and D$^{+}${\textrightarrow} {\ensuremath{\gamma}}K$^{*+}$}",
    eprint = "2410.06500",
    archivePrefix = "arXiv",
    primaryClass = "hep-ex",
    doi = "10.1007/JHEP12(2024)206",
    journal = "JHEP",
    volume = "12",
    pages = "206",
    year = "2024"
}

@article{Aliev:2006vs,
    author = "Aliev, T. M. and Savci, M.",
    title = "{Analysis of the semileptonic $B_c \to B^*_u \ell^{+} \ell^{-}$ decay from QCD sum rules}",
    eprint = "hep-ph/0601267",
    archivePrefix = "arXiv",
    reportNumber = "METU-PHYS-HEP-06-01",
    doi = "10.1140/epjc/s2006-02579-5",
    journal = "Eur. Phys. J. C",
    volume = "47",
    pages = "413--421",
    year = "2006"
}

@article{Melikhov:2000yu,
    author = "Melikhov, D. and Stech, B.",
    title = "{Weak form-factors for heavy meson decays: An Update}",
    eprint = "hep-ph/0001113",
    archivePrefix = "arXiv",
    reportNumber = "HD-THEP-00-01",
    doi = "10.1103/PhysRevD.62.014006",
    journal = "Phys. Rev. D",
    volume = "62",
    pages = "014006",
    year = "2000"
}

@article{Cao:2023gfv,
    author = "Cao, Ye and Zhao, Qiang",
    title = "{Study of weak radiative decays of D0{\textrightarrow}V{\ensuremath{\gamma}}}",
    eprint = "2311.05249",
    archivePrefix = "arXiv",
    primaryClass = "hep-ph",
    doi = "10.1103/PhysRevD.109.093005",
    journal = "Phys. Rev. D",
    volume = "109",
    number = "9",
    pages = "093005",
    year = "2024"
}

@article{Belle:2016mtj,
author = "Abdesselam, A. and others",
    collaboration = "Belle",
    title = "{Observation of $D^0\to \rho^0\gamma$ and search for $CP$ violation in radiative charm decays}",
    eprint = "1603.03257",
    archivePrefix = "arXiv",
    primaryClass = "hep-ex",
    reportNumber = "BELLE-CONF-1601, BELLE-PREPRINT-2016-11",
    doi = "10.1103/PhysRevLett.118.051801",
    journal = "Phys. Rev. Lett.",
    volume = "118",
    number = "5",
    pages = "051801",
    year = "2017"
}

@article{ParticleDataGroup:2026,
    author        = "Takahashi, F. and others",
    collaboration = "Particle Data Group",
    title         = "{Review of Particle Physics}",
    journal       = "Int. J. Mod. Phys. A",
    volume        = "41",
    year          = "2026",
    pages         = "2630011",
    doi           = "10.1142/S0217751X26300115"
}

@article{Workman:2022ynf,
    author         = "Workman, R. L. and others",
    collaboration  = "Particle Data Group",
    title          = "{Review of Particle Physics}",
    journal        = "PTEP",
    volume         = "2022",
    pages          = "083C01",
    year           = "2022",
    doi            = "10.1093/ptep/ptac097",
    note           = "Used for experimental baseline vector meson leptonic decay widths $\Gamma(V \to e^+e^-)$."
}

@article{Aoki:2021kgd,
    author         = "Aoki, Y. and others",
    collaboration  = "Flavour Lattice Averaging Group (FLAG)",
    title          = "{FLAG Review 2021}",
    journal        = "Eur. Phys. J. C",
    volume         = "82",
    number         = "10",
    pages          = "869",
    year           = "2022",
    doi            = "10.1140/epjc/s10052-022-10536-1",
    eprint         = "2111.09849",
    archivePrefix  = "arXiv",
    primaryClass   = "hep-lat",
    note           = "Provides global lattice consensus frameworks for light meson decay constants."
}

@article{Golowich:1994zr,
    author = "Golowich, Eugene and Pakvasa, Sandip",
    title = "{Uncertainties from long range effects in B ---{\ensuremath{>}} K* gamma}",
    eprint = "hep-ph/9408370",
    archivePrefix = "arXiv",
    reportNumber = "UMHEP-411, UH-511-800-94",
    doi = "10.1103/PhysRevD.51.1215",
    journal = "Phys. Rev. D",
    volume = "51",
    pages = "1215--1223",
    year = "1995"
}

@article{Greub:1996wn,
    author = "Greub, Christoph and Hurth, Tobias and Misiak, Mikolaj and Wyler, Daniel",
    title = "{The c ---{\ensuremath{>}} u gamma contribution to weak radiative charm decay}",
    eprint = "hep-ph/9603417",
    archivePrefix = "arXiv",
    reportNumber = "SLAC-PUB-7120, ZU-TH-6-96",
    doi = "10.1016/0370-2693(96)00694-6",
    journal = "Phys. Lett. B",
    volume = "382",
    pages = "415--420",
    year = "1996"
}

@article{Wirbel:1985ji,
    author = "Wirbel, M. and Stech, B. and Bauer, Manfred",
    title = "{Exclusive Semileptonic Decays of Heavy Mesons}",
    reportNumber = "HD-THEP-85-19",
    doi = "10.1007/BF01560299",
    journal = "Z. Phys. C",
    volume = "29",
    pages = "637",
    year = "1985"
}

@article{Charles:1998dr,
    author = "Charles, J. and Le Yaouanc, A. and Oliver, L. and Pene, O. and Raynal, J. C.",
    title = "{Heavy to light form-factors in the heavy mass to large energy limit of QCD}",
    eprint = "hep-ph/9812358",
    archivePrefix = "arXiv",
    reportNumber = "LPTHE-ORSAY-98-77",
    doi = "10.1103/PhysRevD.60.014001",
    journal = "Phys. Rev. D",
    volume = "60",
    pages = "014001",
    year = "1999"
}

@article{Beneke:2000wa,
    author = "Beneke, M. and Feldmann, T.",
    title = "{Symmetry breaking corrections to heavy to light B meson form-factors at large recoil}",
    eprint = "hep-ph/0008255",
    archivePrefix = "arXiv",
    reportNumber = "PITHA-00-20",
    doi = "10.1016/S0550-3213(00)00585-X",
    journal = "Nucl. Phys. B",
    volume = "592",
    pages = "3--34",
    year = "2001"
}

@article{Soares:1995ri,
    author = "Soares, Joao M.",
    title = "{The contribution of the J / Psi resonance to the radiative B decays}",
    eprint = "hep-ph/9503285",
    archivePrefix = "arXiv",
    reportNumber = "TRI-PP-95-6",
    doi = "10.1103/PhysRevD.53.241",
    journal = "Phys. Rev. D",
    volume = "53",
    pages = "241--248",
    year = "1996"
}

@article{Losacco:2024zji,
    author = "Losacco, Nicola",
    title = "{Exclusive c {\textrightarrow} u{\ensuremath{\gamma}} transitions of Bc meson}",
    eprint = "2410.17616",
    archivePrefix = "arXiv",
    primaryClass = "hep-ph",
    reportNumber = "BARI-TH/767-24",
    doi = "10.1051/epjconf/202431400005",
    journal = "EPJ Web Conf.",
    volume = "314",
    pages = "00005",
    year = "2024"
}

@article{Lyon:2012fk,
    author = "Lyon, James and Zwicky, Roman",
    title = "{$A_{\textrm{CP}}[D_{(s)}^{0,+} \to V \gamma]$ from Large ${\cal O}_8$}",
    eprint = "1210.6546",
    archivePrefix = "arXiv",
    primaryClass = "hep-ph",
    reportNumber = "EDINBURGH-12-16, CP{\textasciicircum}3-ORIGINS-2012-23, DIAS-2012-24",
    doi = "10.1103/PhysRevD.106.053001",
    journal = "Phys. Rev. D",
    volume = "106",
    number = "5",
    pages = "053001",
    year = "2022"
}

@article{Gisbert:2020vjx,
    author = "Gisbert, Hector and Golz, Marcel and Mitzel, Dominik Stefan",
    title = "{Theoretical and experimental status of rare charm decays}",
    eprint = "2011.09478",
    archivePrefix = "arXiv",
    primaryClass = "hep-ph",
    reportNumber = "DO-TH 20/13, CERN-OPEN-2020-012",
    doi = "10.1142/S0217732321300020",
    journal = "Mod. Phys. Lett. A",
    volume = "36",
    number = "04",
    pages = "2130002",
    year = "2021"
}

@article{Biswas:2017eyn,
    author = "Biswas, Aritra and Mandal, Sanjoy and Sinha, Nita",
    title = "{Searching for New physics in Charm Radiative decays}",
    eprint = "1702.05059",
    archivePrefix = "arXiv",
    primaryClass = "hep-ph",
    doi = "10.1142/S0217751X18501944",
    journal = "Int. J. Mod. Phys. A",
    volume = "33",
    number = "32",
    pages = "1850194",
    year = "2018"
}

@article{Colquhoun:2015BcDecayConstants,
 author = {Colquhoun, B. and others},
 collaboration = {HPQCD},
 title = {{$B$-meson decay constants: a more complete picture from full lattice QCD}},
 journal = {Phys. Rev. D}, volume = {91}, pages = {114509}, year = {2015},
 doi = {10.1103/PhysRevD.91.114509},
 eprint = {1503.05762}, archivePrefix = {arXiv}, primaryClass = {hep-lat}
}

@article{Bazavov:2018BDecayConstants,
 author = {Bazavov, A. and others},
 collaboration = {Fermilab Lattice and MILC},
 title = {{$B$- and $D$-meson leptonic decay constants from four-flavor lattice QCD}},
 journal = {Phys. Rev. D}, volume = {98}, pages = {074512}, year = {2018},
 doi = {10.1103/PhysRevD.98.074512},
 eprint = {1712.09262}, archivePrefix = {arXiv}, primaryClass = {hep-lat}
}

@article{Cai:2026BDecayConstants,
 author = {Cai, Mengchu and Du, Hai-Yang and Jiang, Xiangyu and Sun, Peng and Sun, Wei and Wang, Ji-Hao and Yang, Yi-Bo},
 collaboration = {CLQCD},
 title = {{Accurate B meson and Bottomonium masses and decay constants from the tadpole improved clover ensembles}},
 year = {2026}, eprint = {2603.01846}, archivePrefix = {arXiv}, primaryClass = {hep-lat}
}

@article{CostaiReina:2026BcDecayConstants,
 author = {{Costa i Reina}, Eduard and Gubernari, Nico and W{\"u}thrich, Zachary},
 title = {{Decay constants of the $B_c$ mesons with vector and tensor currents}},
 journal = {Eur. Phys. J. C}, volume = {86}, pages = {1204}, year = {2026},
 doi = {10.1140/epjc/s10052-026-16381-w},
 eprint = {2606.25733}, archivePrefix = {arXiv}, primaryClass = {hep-ph}
}

@article{ATLAS:2026BcStarObservation,
 author = {Aad, G. and others}, collaboration = {ATLAS},
 title = {{Observation of a $B_c^{*+}$ meson with the ATLAS detector}},
 year = {2026}, eprint = {2605.16228}, archivePrefix = {arXiv}, primaryClass = {hep-ex}
}

@article{Meng:2024DsStarRadiative,
 author = {Meng, Yu and Dang, Jin-Long and Liu, Chuan and Liu, Zhaofeng and Shen, Tinghong and Yan, Haobo and Zhang, Ke-Long},
 title = {{Lattice QCD calculation of the $D_s^*$ radiative decay with (2+1)-flavor Wilson-clover ensembles}},
 journal = {Phys. Rev. D}, volume = {109}, pages = {074511}, year = {2024},
 doi = {10.1103/PhysRevD.109.074511},
 eprint = {2401.13475}, archivePrefix = {arXiv}, primaryClass = {hep-lat}
}

@article{Donald:2014DsStarWidth,
 author = {Donald, G. C. and Davies, C. T. H. and Koponen, J. and Lepage, G. P.},
 title = {{Prediction of the $D_s^*$ width from a calculation of its radiative decay in full lattice QCD}},
 journal = {Phys. Rev. Lett.}, volume = {112}, pages = {212002}, year = {2014},
 doi = {10.1103/PhysRevLett.112.212002},
 eprint = {1312.5264}, archivePrefix = {arXiv}, primaryClass = {hep-lat}
}

@article{PDG:2025KstarInputs,
 author = {Navas, S. and others}, collaboration = {Particle Data Group},
 title = {{Review of Particle Physics}},
 journal = {Phys. Rev. D}, volume = {110}, pages = {030001}, year = {2024},
 doi = {10.1103/PhysRevD.110.030001},
 note = {2025 update; $K^*(892)$ listing and charged-pseudoscalar decay constants},
 url = {https://pdg.lbl.gov/2025/listings/rpp2025-list-K-star-892.pdf}
}

@article{Feldmann:1998vh,
 author = {Feldmann, Thorsten and Kroll, Peter and Stech, Berthold},
 title = {Mixing and Decay Constants of Pseudoscalar Mesons},
 journal = {Phys. Rev. D}, volume = {58}, pages = {114006}, year = {1998},
 doi = {10.1103/PhysRevD.58.114006}, eprint = {hep-ph/9802409}, archivePrefix = {arXiv}
}

@article{Cai:2026CLQCD,
  author        = "Cai, Mengchu and Du, Hai-Yang and Jiang, Xiangyu
                   and Sun, Peng and Sun, Wei and Wang, Ji-Hao
                   and Yang, Yi-Bo",
  collaboration = "CLQCD",
  title         = "{Accurate B meson and Bottomonium masses and decay
                    constants from the tadpole improved clover ensembles}",
  eprint        = "2603.01846",
  archivePrefix = "arXiv",
  primaryClass  = "hep-lat",
  year          = "2026"
}

@article{Wang:2025voj,
    author = "Wang, Zhi Jun and Gao, Di and Zhang, Kai Kai and Ma, Yuan Yuan and Sun, Yan Jun",
    title = "{Radiative decays of vector mesons with light-cone sum rules}",
    eprint = "2511.12996",
    archivePrefix = "arXiv",
    primaryClass = "hep-ph",
    doi = "10.1103/nm8j-l3ky",
    journal = "Phys. Rev. D",
    volume = "113",
    number = "11",
    pages = "116003",
    year = "2026"
}

@article{Ball:2004rg,
    author = "Ball, Patricia and Zwicky, Roman",
    title = "{$B_{d,s} \to  \rho, \omega, K^*, \phi$ decay form-factors from light-cone sum rules revisited}",
    eprint = "hep-ph/0412079",
    archivePrefix = "arXiv",
    reportNumber = "IPPP-04-74, DCPT-04-48, TPI-MINN-04-39",
    doi = "10.1103/PhysRevD.71.014029",
    journal = "Phys. Rev. D",
    volume = "71",
    pages = "014029",
    year = "2005"
}

@article{Pullin:2021ebn,
    author = "Pullin, Ben and Zwicky, Roman",
    title = "{Radiative decays of heavy-light mesons and the $ {f}_{H,{H}^{\ast },{H}_1}^{(T)} $ decay constants}",
    eprint = "2106.13617",
    archivePrefix = "arXiv",
    primaryClass = "hep-ph",
    reportNumber = "CP3-Origins-2020-13 DNRF90",
    doi = "10.1007/JHEP09(2021)023",
    journal = "JHEP",
    volume = "09",
    pages = "023",
    year = "2021"
}
\end{document}